\documentclass[aps,prb,showpacs,twocolumn,eqsecnum,floatfix]{revtex4-1}
\usepackage[utf8]{inputenc}
\usepackage{amsmath,amssymb,subfigure,psfrag}
\usepackage{graphicx}
\usepackage{bm}
\begin{document}
\title{Quantum Entanglement and Thermal Reduced Density Matrices in Fermion and Spin Systems on Ladders}
\author{Xiao Chen}

\author{Eduardo Fradkin}
\affiliation{Department of Physics  and Institute for Theoretical Condensed Matter Theory, University of Illinois at Urbana-Champaign, 1110 West Green Street, Urbana,
Illinois 61801-3080, USA}

\date{\today}
\begin{abstract}
Numerical studies of the  reduced density matrix of a gapped spin-1/2 Heisenberg antiferromagnet on a two-leg ladder find that it has the same form as the Gibbs density matrix of a gapless spin-1/2 
Heisenberg antiferromagnetic chain at a finite temperature determined by the spin gap of the ladder. We investigate this interesting result by considering a model of  free 
fermions on a two-leg ladder (gapped by the inter-chain tunneling operator)  and in spin systems on a ladder with a gapped ground state using exact solutions and several controlled approximations. 
We calculate the reduced density matrix and the entanglement entropy for a leg of the ladder (i.e.  cut made between the chains).   In the fermionic system we find the exact form of the reduced 
density matrix for one of the chains and determine the entanglement spectrum explicitly. Here we find that in the weak tunneling limit of the ladder the entanglement entropy of one chain of the gapped ladder has a simple and universal form dictated by conformal invariance.  In the case of the spin system, we consider the strong coupling limit by using perturbation theory and get 
the reduced density matrix by the Schmidt decomposition. The entanglement entropies of a general gapped system of two coupled conformal field theories (in 1+1 dimensions)  is discussed using the replica trick and scaling arguments. 
We  show that 1) for a system with a bulk gap  the reduced density matrix has the form of a thermal density matrix, 2) the long-wavelength modes of one subsystem (a chain) of a gapped coupled system are always thermal, 3) the von Neumann entropy equals to the thermodynamic entropy of one chain, 
and 4) the bulk gap plays the role of effective temperature. 
\end{abstract} 

\maketitle
\section{Introduction}
\label{sec:introduction}

The thermodynamic entropy is a key concept in Statistical Mechanics, and measures disorder and randomness in a macroscopic system. 
The thermodynamic entropy $S_T$  for a quantum system in thermal equilibrium at temperature $T$  is defined as 
\begin{equation}
S_T\equiv-\textrm{Tr}\rho_T \ln \rho_T
\end{equation}
 where 
\begin{equation}
\rho_T\equiv \frac{1}{Z} e^{-\beta H}
\end{equation} 
is (thermal density) matrix, of the Gibbs ensemble at temperature $T$, and $Z\equiv \textrm{Tr} \exp(-\beta H)$ is the Gibbs partition function with $\beta=1/T$. 
At zero temperature, the system is in its ground state 
(with an at most finite degeneracy)  and in this limit the thermodynamic entropy vanishes, $S_T=0$. 

On the other hand, the entanglement entropy is a measure of the non-local correlations  of a pure quantum state. 
The entanglement entropy for subsystem $A$  is a measure of the quantum entanglement between $A$ and $B$ (and viceversa). and it is defined as follows.  
Let us consider an extended  system in a pure state $|\Psi\rangle$, and define a partition of system into two subsystems, $A$ and $B$ with 
common boundary $\Gamma=\partial A=\partial B$. We will denote by 
\begin{equation}
\rho_{A \cup B}=|\Psi\rangle \langle \Psi |
\end{equation}
 the density matrix of the pure state $|\Psi\rangle$, and by 
 \begin{equation}
 \rho_A=\textrm{Tr}_B \rho_{A \cup B}, \quad \rho_B=\textrm{Tr}_A \rho_{A \cup B}, 
 \end{equation}
  the (normalized) reduced density matrices of the subsystems $A$ and $B$ of the partition (which satisfy $\textrm{Tr} \rho_A=\textrm{Tr} \rho_B=1$).  
  Then, the (von Neumann)  entanglement entropy is  given by 
  \begin{equation}
  S_{vN}(A)\equiv- \textrm{Tr} \rho_A \ln\rho_A
  \label{von-neumann}
  \end{equation}
  Since the full system $A \cup B$ is in a pure state, $|\Psi \rangle$, the von Neumann entanglement entropy is the same for both members of the partition, $S_{vN}(A)=S_{vN}(B)$. 
  Similarly, the R\'enyi entropies $S_n$ are given by ($n>1$)
  \begin{equation}
  S_n=\frac{1}{1-n} \ln \textrm{Tr} \rho_A^n
  \label{eq:Renyi}
  \end{equation}
  and are also symmetric under the exchange of regions $A$ and $B$. The von Neumann and R\'enyi entropies are related by the ``replica trick'' formula\cite{Callan1994,Holzhey1994,Calabrese2004}
  \begin{equation}
  \label{eq:replica-trick}
  \end{equation}
  which is understood as an analytic continuation.

The behavior of the entanglement entropy has been the focus of study in several areas of physics. A particular focus of interest has been the scaling of the 
entanglement entropy with the linear size $\ell$ of the subsystem, assumed to be much smaller than the linear size $L$ of the system as a whole, 
$\ell \ll L$. It is known that for a generic state in spatial dimension $d$  the entanglement entropy scales with the area of the subsystem\cite{Bombelli1986,Srednicki1993,Eisert2010}  
$S_{vN}(\ell)=\alpha \ell^{d-1}$ where $\alpha$ is a non-universal constant determined by the short-distance correlations of the wave function. 
This result is reminiscent of the area law of the entropy of black holes\cite{Bekenstein1973,Hawking1975} where the constant is instead determined by the Planck scale. 
Of particular interest is the fact that quantum entanglement also encodes universal information of the non-local correlations  of the many-body wavefunction of the macroscopic 
quantum system.\cite{Calabrese2004,Kitaev2006,Levin2006,Fradkin2006,Wen2012}

Although the von Neumann {\it entanglement} entropy $S_{vN}$ has the same formal definition as the {\it thermodynamic} entropy $S_T$, these are conceptually different quantities. 
In this paper we will be interested  in under what circumstances can the {\em reduced density matrix} $\rho_A$ of a subsystem of a system in its {\em ground state} $|\Psi\rangle$ 
define an effective Gibbs ensemble for the subsystem at some effective temperature $T_{\rm eff}$. For this equivalence to be meaningful it should be possible to express  
the reduced density matrix, whose spectrum is by definition non-negative,  in terms of an effective {\em local}  so-called entanglement Hamiltonian, that we will denote by  $H_E$, whose spectrum is the entanglement spectrum.\cite{Li2008} 
If this equivalence holds, then the reduced density matrix takes the thermal form
\begin{equation}
\rho_A=\frac{1}{Z_{\rm eff}} e^{-\beta_{\rm eff} H_E}
\end{equation}
where $\beta_{\rm eff}=1/T_{\rm eff}$, and the normalization factor $Z_{\rm eff}$ plays the role of a an effective partition function.
Since the reduced density matrix is, by definition, a Hermitian matrix it is obvious that a suitable Hermitian operator $H_E$ can always be defined. However it is not obvious, and in general it is not true, 
that $H_E$ should also be local and, even more, what connection it may bear, if any, with the Hamiltonian $H$ of the combined quantum system of which the state $|\Psi\rangle$ is its ground state or with the Hamiltonian $H_A$ of subsystem $A$ (and similarly with $B$).

In this paper we will consider systems made of two identical subsystems which are coupled to each other in the bulk. In this case, both subsystems are thermodynamically large and 
neither can be regarded as a ``heat bath'' for the other. Here we will focus on the special (and interesting) problem in which the two identical subsystems are one-dimensional and are 
separately at quantum criticality. The problem that we want to address is under what circumstances can the reduced density matrix of one of these subsystems have a Gibbs form at some effective 
temperature $T_{\rm eff}$ with a local (and Hermitian) entanglement Hamiltonian $H_E$.  We are motivated by some recent numerical results by Poilblanc\cite{Poilblanc2010} who used an exact 
diagonalization technique to determine the entanglement Hamiltonian for one leg of spin-1/2 quantum antiferromagnet on a two-leg ladder. 
Over some range of values of the inter-leg exchange interaction, 
Poilblanc found that the reduced density matrix of one leg is the same as the thermal (Gibbs) density matrix of a spin-1/2 quantum Heisenberg {\em chain} at an effective temperature 
(determined by the spin gap of the ladder). Similar results have also been found in other fully gapped systems such as AKLT models on ladders\cite{Katsura2010,Lou2011} and in the entanglement of spin and orbital degrees of freedom in Kugel-Khomski models in one dimension.\cite{Lundgren2012} To this end we examine this question first in  an exactly solvable system of free fermions on a ladder, with a gapped ground state. 
Next we examine the same problem in the spin-1/2 ladder in the strong inter-leg coupling regime, a system recently discussed also by La\"uchli and Schliemann\cite{Lauchli2012} and by 
Qi, Katsura and Ludwig\cite{Qi2012} in 2D topological phases. Next we formulate a scaling hypothesis for the entanglement entropy in the weak coupling limit, 
where the combined system 
can be regarded as  being a perturbed conformal field theory, and test its validity in the free-fermion system. 
We finally compare with results in a system of two coupled Luttinger liquids in a gapless combined ground state\cite{Furukawa2011}. 

An important question is whether the effective entanglement Hamiltonian $H_E$ is local and what is its relation with the (local) Hamiltonian of the decoupled subsystems. 
We will see below that if we insist that the entanglement Hamiltonian $H_E$  be fully local (i.e. at the scale of the lattice spacing) the energy gap of the coupled system 
 has to be much larger than the coupling constants (end hence the energy scales) of the subsystems. 
 In this regime all the degrees of freedom of the subsystem are thermal. 
 However we will see in an explicitly solvable free-fermion example that in regimes in which the gap is small (compared with other scales of the problem), 
  the reduced density matrix  for the {\em long wavelength} degrees of freedom of subsystem $A$ is  thermal  with a {\em local} effective {\em continuum} effective entanglement Hamiltonian 
  which is the same as the Hamiltonian $H_A$ of the  low-energy conformal field theory of the decoupled subsystem $A$.
  Moreover, in this regime the structure of the effective long-wavelength entanglement entropy has a  form which is determined entirely by conformal invariance. 
  This observation leads us to conjecture that this result and not a peculiarity of the free fermion system but it is actually a general property of gapped systems of this type.
  The separation of the entanglement spectrum into a long-wavelength universal (and thermal)  piece and a short-distance non-universal piece that we found 
  in this free-fermion model is 
  in line with what was found  by Li and Haldane.\cite{Li2008}  These authors showed that the low-(pseudo)energy modes of the entanglement 
  Hamiltonian of  fractional quantum Hall fluids of a two-dimensional electron gas have the same universal structure as the low-energy states of the edge states of the 
  same fractional quantum Hall state (on a disk geometry).

 We should note that the question we are asking here is conceptually different from the central axiom of Statistical Mechanics stating 
 that a subsystem weakly coupled to a much larger system (the ``heat bath'') 
 can reach thermal equilibrium at a temperature determined by the larger system. It is an axiom of Statistical Mechanics that the equilibrium 
 state of the subsystem is in the Gibbs Ensemble, 
 and that this equilibrium state is universally reached irrespective of the specific dynamics.\cite{Schrodinger1952} 
It is known rigorously that the reduced density matrix of a subsystem has a Gibbs form if the total system is in a ``typical state'', {\it i.e.} a state drawn from some statistical ensemble, 
which is assumed to be a typical state of the spectrum of the full (and generic) Hamiltonian $H$.\cite{Tasaki1998,Goldstein2006,Popescu2006}
 However the ground state of the Hamiltonian is hardly a typical state and one generally does not expect to find a Gibbssian reduced density unless the ground state has special properties.

The paper is organized as follows. In Section \ref{sec:free-fermion} we consider a system of free fermions on a  ladder which is gapped by the inter-chain tunneling amplitude. 
This problem is exactly solvable and the reduced density matrix can be determined explicitly.\cite{Peschel2003,Peschel2009,Peschel2011} In Section \ref{sec:strong-coupling} we consider the spin ladder problem in the strong inter-chain coupling limit and we show that in this limit the 
reduced density matrix is that of a spin-1/2 quantum Heisenberg chain.  In Section \ref{sec:weak-coupling} 
we use the insights obtained in the free-fermion system of Section \ref{sec:free-fermion} to formulate a scaling ansatz for the form of the entanglement entropy for a system of two weakly coupled quantum critical systems 
(which can be regarded as a perturbed conformal field theory). Here we conjecture a general form of the 
the scaling behavior of the entanglement entropies, and infer that
the reduced density matrix for the low energy degrees of freedom of the subsystem is the thermal Gibbs density matrix of the conformal field theory at a 
temperature determined by the gap scale. In Section \ref{sec:coupled-LL} we consider the case of two coupled Luttinger liquids with a joint gapless ground state. 
Our conclusions are summarized in Section \ref{sec:conclusions}.

\section{A Free Fermion Model}
\label{sec:free-fermion}

In this section we will consider a two-leg ladder model of free fermions which are gapped by the inter-chain tunneling. 
Consider a two-leg ladder model with Hamiltonian \cite{Jaefari2012}
\begin{align}
H=&-t\sum_j(e^{i\Phi/2}c_{A,j+1}^{\dag}c_{A,j}+e^{-i\Phi/2}c_{B,j+1}^{\dag}c_{B,j}+h.c.)\nonumber\\
&+t_{\bot}\sum_j(c_{A,j}^{\dag}c_{B,j}+c_{B,j}^{\dag}c_{A,j})
\label{eq:HladderPhi}
\end{align} 
in which $c_{A,j}$ and $c_{B,j}$ are the fermion operators on chain $A$ and on chain $B$, respectively. Here $t$ is the hopping amplitude 
along the chains and $t_{\bot}$ is the hopping amplitude along the 
rungs (between the chains). For each plaquette of the ladder there is a flux  $\Phi$ introduced in the Hamiltonian through minimal coupling 
(the Peierls substitution). As the flux per plaquette $\Phi$ varies from $0$ to $\pi$ the spectrum evolves continuously from a regime with two gapless branches 
(for $\Phi \sim 0$) to a fully gapped spectrum (for $\Phi \sim \pi$). 
Since in this paper we are interested in the gapped case,  we consider only the simple case in 
which the flux per plaquette is half of the flux quantum and hence  $\Phi=\pi$ (in units in which $\hbar=c=e=1$). 
The behavior of entanglement in the gapless regime is similar to what is discussed in Section \ref{sec:coupled-LL}.

In momentum space, the Hamiltonian of this model becomes:
\begin{equation}
H=\int_{-\pi}^{\pi}\frac{dk}{2\pi} \mathcal{H}(k)
\label{eq:Hpi}
\end{equation}
For flux $\Phi=\pi$, the Hamiltonian $\mathcal{H}(k)$ of Eq.\eqref{eq:Hpi}  is 
\begin{align}
\mathcal{H}(k)=&2t \sin k \; \left( c_{A}(k)^{\dag}c_{A}(k)-  c_{B}(k)^{\dag}c_{B}(k)\right) \nonumber\\
&-t_{\bot}\left(c_{A}(k)^{\dag}c_{B}(k)+c_{B}(k)^{\dag}c_{A}(k)\right)
\label{eq:mathcalHk}
\end{align}
The Hamiltonian can be diagonalized by a change of basis
\begin{align}
c_{A}(k)=&\cos \Big(\frac{\xi(k)}{2}\Big)c^a(k)-\sin \Big(\frac{\xi(k)}{2}\Big)c^b(k)\nonumber \\
c_{B}(k)=&\sin \Big(\frac{\xi(k)}{2}\Big)c^a(k)+\cos \Big(\frac{\xi(k)}{2}\Big)c^b(k)
\end{align}
where $b$ and $a$ label the bonding and the anti-bonding bands of the ladder, respectively, and
$\xi(k)$ is defined as
\begin{align}
\sin \Big(\frac{\xi(k)}{2}\Big)=&\frac{u(k)}{\sqrt{1+u^2(k)}}\nonumber \\
\cos \Big(\frac{\xi(k)}{2}\Big)=&\frac{1}{\sqrt{1+u^2(k)}}
\end{align}
where $u(k)$ is given by
\begin{equation}
t_{\bot}u(k)=2t \sin k+\sqrt{(2t \sin k)^2+t_{\bot}^2}
\label{eq:uofk}
\end{equation}
The dispersion relations for the bonding and anti-bonding bands are
\begin{equation}
E(k)=\pm\sqrt{(2t \sin k)^2+t_{\bot}^2}
\label{eq:dispersion}
\end{equation}

At half filling, the bonding band is filled and the anti-bonding band is empty. As can be seen from Eq.\eqref{eq:dispersion}, 
the excitation energy $E(k)$ is smallest at $k=0,\pi$, where the spectrum has an energy gap of $2t_\bot$. 

As in all fermionic systems in 1D, this system can also be put in the form of 1D Dirac fermions with two two-component spinor fields, with the components being the right and left moving amplitudes 
near the two Fermi points at $k=0,\pi$. Therefore the low-energy degrees of freedom of this ladder (with flux $\Phi=\pi$) are described by two species (bonding and anti-bonding) 
Dirac spinors each with velocity $v=2t$ and mass gap $mv^2=t_\bot$. 
This can be done, more formally, by combining the right moving fermion from the $A$ chain (with $k \sim 0$), $R_A(k)$ and the left-moving fermion from the $B$ chain 
(also with $k\sim 0$), $L_B(k)$, into a two-component (Weyl) spinor. Similarly a second spinor 
 can be constructed where $\tilde R_B(k)$ is the right-moving component of the fermion on the $B$ chain with momentum 
$-\pi+k$ and $\tilde L_A(k)$ is the left-moving fermion from the $A$ chain with momentum $\pi-k$. 
\begin{equation}
\psi_1(k)=
\begin{pmatrix}
R_A(k\\
L_B(k)
\end{pmatrix}, 
\qquad 
\psi_2(k)=
\begin{pmatrix}
\tilde R_B(k)\\
\tilde L_A(k)
\end{pmatrix}
\label{eq:spinors}
\end{equation}
The tunneling matrix element $t_\bot$, which mixes right and left movers with the same momenta on both chains,  opens the (same) mass gap $m \propto t_\bot$ in both 
Dirac spinors. The effective (continuum) low energy Hamiltonian density for this system is 
\begin{equation}
\mathcal{H}= \sum_{a=1,2}\left(\psi_a^\dagger  \sigma_3 i v \partial_x \psi_a +mv^2 \psi^\dagger_a \sigma_1 \psi_a\right)
\end{equation}
where $a=1,2$ labels the two spinors and $\sigma_1$ and $\sigma_3$ are the two $2 \times 2$ Pauli matrices (which act on the components of each spinor).

Below we will calculate the reduced density matrix for chain $A$ by making a cut between the chains and trace out chain $B$. We can now use the results of  Peschel \cite{Peschel2003} 
for free-fermion system to find the entanglement Hamiltonian 
\begin{equation}
\widetilde H_E\equiv -\ln \rho_A
\label{eq:tildeH}
\end{equation}
 for subsystem $A$, which has the explicit form 
\begin{equation}
\widetilde H_E=\sum_{i,j=1}^N \widetilde H_{ij}c^{\dag}_i c_j
\label{eq:peschel}
\end{equation}
where the matrix $\widetilde H_{ij}$ takes the form
\begin{equation}
\widetilde H_{ij}=\left(\ln\left[(C^{-1}-1)\right]\right)_{ij}
\label{eq:Hij}
\end{equation}
The creation and annihilation operators $c_I^\dagger$ and $c_i$ in Eq.\eqref{eq:peschel} are labelled by the sites of subsystem $A$.
In Eq.\eqref{eq:Hij} $C_{ij}$ is the correlation function matrix (the fermion propagator at equal times) whose  matrix elements in momentum space are 
\begin{align}
C_{kk^{\prime}}=& \langle c^\dagger_{A}(k) c_{A}(k^{\prime})\rangle =2\pi \delta(k-k^{\prime}) \sin ^2\Big(\frac{\xi(k)}{2}\Big)\nonumber\\
=&2\pi \delta(k-k^{\prime}) \frac{u^2(k)}{1+u^2(k)}
\label{eq:eq-time}
\end{align}

Combining the above two equations, we find that the entanglement Hamiltonian (in momentum space) has the standard  form
\begin{equation}
\widetilde H_E=\int_{-\pi}^\pi \frac{dk}{2\pi} \; \omega(k) \; c^{\dag}(k)c(k)
\label{eq:H_E-fermion-ladder}
\end{equation}
where 
\begin{equation}
\omega(k)= \ln u^2(k)
\end{equation}

By inspection of Eq.\eqref{eq:uofk} we see that as $k \to 0$ the quantity $u(k) \simeq 1 + v k/t_\bot+ O(k^2)$, and similarly as $k \to \pi$. 
Thus the one-particle spectrum $\omega(k)$ vanishes linearly as $k \to 0,\pi$. In other terms, the long-wavelength modes (with $k\sim 0, \pi$)  
of the one-particle entanglement spectrum is that of a system of massless fermions, $\omega(k) \simeq 2 v k/t_\bot$, with the modes near $k=0$ representing right-movers 
and the modes near $k=\pi$ representing left movers, respectively.

If we define the inverse temperature 
$\beta_{\rm eff}=(t_{\bot}/2)^{-1}$, we can rewrite the reduced density matrix $\rho_A$ as
\begin{equation}
\rho_A=\rho_{T_{\rm eff}}=\frac{1}{Z} e^{-\beta_{\rm eff} H_A}
\end{equation}
We can see that $\rho_A$ has the same form as $\rho_T$ for chain A with $T_{\rm eff}=t_\bot/2$ playing the role of the temperature.
Therefore,  the entanglement Hamiltonian for the long-wavelength modes (near $k=0$ and $k=\pi$,  always has the form (regardless of the strength of the tunneling amplitude $t_{\bot}$)
\begin{align}
\widetilde H_E=&\int_{-\Lambda}^\Lambda \frac{dk}{2\pi} \; \frac{4ta}{t_{\bot}}k \big(R^\dagger(k)R(k)-L^\dagger(k) L(k)\big) \nonumber\\
=&\beta_{\rm eff} H_A
\label{eq:H_E-long-wavelengths}
\end{align}
where $R(k)$ represent the right-moving modes (with $k\sim 0$) and $L(k)$ the left-moving modes (with wave vector $\pi-k$), respectively, $v=2ta$ is the velocity of the modes,
and $\Lambda \sim \pi/a$ is a momentum cutoff (and $a$ is the lattice spacing which we have set to $1$). In other terms, the long-wavelength entanglement Hamiltonian for chain 
$A$ is the same as the low-energy Hamiltonian $H_A$ for the Dirac fermions of the decoupled chain.
Therefore the long-wavelength  reduced density 
matrix is the Gibbs density matrix of a system of a massless Dirac fermion (with velocity $2t$) at temperature $T_{\rm eff}=\beta_{\rm eff}^{-1}=t_\bot/2$.

On the other hand, in the strong coupling (tunneling)  limit $t_{\bot}\gg t$, in which there is a large energy gap in the spectrum of the fermions,  
the entanglement Hamiltonian has the simple form
\begin{align}
\widetilde H_E=&\int_{-\pi}^\pi \frac{dk}{2\pi} \; \frac{4t}{t_{\bot}}\sin k \; c^{\dag}(k)c(k)\nonumber\\
=&\beta_{\rm eff}\; i t \sum_{n=1}^N \; c(n)^\dagger c(n+1)+\textrm{h.c.}
\label{eq:H_E-strong-tunneling}
\end{align}
In this limit the reduced density matrix for leg $A$ is that of a single chain of free fermions with  hopping amplitude $it$ 
at an effective temperature  $T_{\rm eff}=t_\bot/2$. 

This effective temperature is in fact much higher than the bandwidth $4t$ of the fermionic spectrum of the chain. 
Thus, the statistical ensemble of the chain defined by the strong tunneling limit is 
essentially the classical Gibbs ensemble.
In the strong tunneling limit the entanglement Hamiltonian is a local operator of the chain degrees of freedom. 
Clearly the corrections to this strong tunneling limit lead to an effective entanglement 
Hamiltonian which becomes increasingly non-local. Nevertheless, these apparently non-local lattice 
operators only contribute with irrelevant operators in the long-wavelength regime. 

In summary, in this free fermion ladder model the reduced density matrix of a chain has a Gibbs form with an effective entanglement Hamiltonian $H_E$, given in Eq.\eqref{eq:H_E-fermion-ladder}. 
Since the reduced density matrix for chain $A$ is thermal, the von Neumann entanglement entropy is equal to the thermodynamic entropy of the 1D quantum system described by the Hamiltonian 
$H_E$. As the strength of the tunneling matrix element $t_\bot$ increases, the fraction of the entanglement spectrum that is thermal also increases ranging from only the
long-wavelength modes of chain $A$  for $t_\bot \ll t$ to all of the modes for $t_\bot \gg t$, with an effective temperature $T_{\rm eff}=t_\bot/2$. 
Nevertheless, the long-wavelength modes, i.e. the lowest eigenvalues of the entanglement Hamiltonian,  which are always thermal, have universal properties.

The free energy of a system of 1D massless Dirac fermions (in a system of length $L$ in the thermodynamic limit)  at temperature $T$ is that of a conformal field theory 
with central charge $c=1$ (see Refs. [\onlinecite{Affleck1986,Blote1986}])
\begin{equation}
F=-T \ln Z=\varepsilon_0 L -\frac{\pi c}{6v} T^2L
\label{eq:1D-fermion-free-energy}
\end{equation}
where $\varepsilon_0$ is the (non-universal) ground state energy density, $c$ is the central charge of the conformal field theory and $v$ is the velocity of the modes. 
The last term in Eq.\eqref{eq:1D-fermion-free-energy} is universal and it is well known low temperature (the Casimir term) contribution to the free energy of a conformal field theory.  
The form of this term is determined by the conformal anomaly of the conformal field theory.\cite{Affleck1986,Blote1986}
From here it follows that the thermodynamic entropy $S_T$ of this 1D quantum critical system is
\begin{equation}
S_T=-\frac{\partial F}{\partial T}=\frac{\pi c}{3v} T L
\label{eq:1D-cft-entropy}
\end{equation}
Depending on the boundary conditions, the entropy $S_T$ may have the finite limiting value $\ln g$ as $T \to 0$, where $g$ is a universal number that depends on the boundary conditions and can be interpreted as a ground state ``degeneracy'' (even though it is generally not an integer).\cite{Affleck1991}

Since we have shown that the reduced density matrix of a leg of  the fermion ladder with flux $\Phi=\pi$ per plaquette is, in the long wavelength limit, identical to the Gibbs 
density matrix of a system of massless Dirac fermions in 1D, we can apply the above results from CFT to the present case.
 It then follows that the thermodynamic entropy of a system of 1D massless Dirac fermions at finite temperature $T$ is the same as the von Neumann entanglement entropy 
 $S_{vN}$ of  chain $A$ (also in the long wavelength limit) with a temperature $T=T_{\rm eff}=M$, given by the mass gap of the fermion ladder.

It is an elementary excercise to compute the R\'enyi entropies, $S_n$.  Indeed in this system the trace of the $n$th power of the (unnormalized) reduced density matrix 
$\rho_A$ is now equal to the partition function of a free Dirac fermion at temperature $T_{\rm eff}/n$,
\begin{align}
\textrm{Tr} \rho_A^n=&Z_F\left(T=\frac{T_{\rm eff}}{n}\right)\nonumber\\
=& \exp\left(-\frac{n}{T_{\rm eff}} \varepsilon_0 L+\frac{\pi c}{6v} \frac{T_{\rm eff}}{n} L\right)
\end{align}
where we have purposely left the explicit dependence on the central charge $c$ of the CFT (although for the free fermion case $c=1$). 
We will return to this expression in Section \ref{sec:weak-coupling}. 
Therefore
\begin{equation}
\frac{\textrm{Tr} \rho_A^n}{\left(\textrm{Tr} \rho_A\right)^n}\equiv \textrm{Tr} \hat \rho_A^n=\exp\left[\frac{\pi c}{6v} \left(\frac{1}{n}-n\right) T_{\rm eff} L\right]
\end{equation}
where we denoted by $\hat \rho_A$ the normalized reduced density matrix, i.e. $\textrm{Tr} \hat \rho_A=1$.
From here we find that the von Neumann entanglement entropy $S_{vN}$ is given by
\begin{equation}
S_{vN}=-\frac{\partial}{\partial n}\textrm{Tr} \hat \rho_A^n\Big|_{n\to 1}
=\frac{\pi c}{3v} T_{\rm eff} L=S(T_{\rm eff})
\end{equation}
which agrees with the thermodynamic entropy $S(T_{\rm eff})$ at temperature $T_{\rm eff}$ (as it should). Similarly, the R\'enyi entropy $S_n$ is given by the result (valid for $n>1$)
\begin{equation}
S_n= \frac{\pi c}{6v} \left(1+\frac{1}{n}\right) T_{\rm eff} L
\end{equation}
Notice that $S_1=\lim_{n \to 1} S_n= S_{vN}$ as it should.

We close this section with a comment of the correlators. The equal-time fermionic correlators, i.e. the equal time propagators (or Green functions), of a theory of massive fermions in $1+1$ dimensions has an asymptotic exponential decay $\exp (-m|x|)$ (where $m$ is the mass gap) with a power law correction prefactor $(m|x|)^{-1/2}$. This behavior is correctly reproduced by Eq.\eqref{eq:eq-time} (as it should). However the equal-time correlation function of gapless Dirac fermions at temperature $T$ has a pure exponential decay of the form $\pi T/\sinh(2\pi T |x|)$, which does not have a power law prefactor correction. This apparent difference is the result of the long-wavelength approximation used in the entanglement Hamiltonian of Eq.\eqref{eq:H_E-long-wavelengths}. 

\section{Entanglement in Strongly Coupled Systems}
\label{sec:strong-coupling}

For a free fermion model, such as the one discussed in Section \ref{sec:free-fermion}, we can exactly calculate the 
reduced density matrix for any value of the coupling constant.
For the general case of two arbitrary coupled systems which are not free the computation of the reduced density matrix is non-trivial.
However, in the strong coupling limit we can still use the perturbation theory to calculate the reduced density matrix. This is the approach we will follow here. 
A similar calculation was done by La\"uchli and Schliemann.\cite{Lauchli2012} 

In general the Hamiltonian will have the form $H=H_0+H_{\rm pert}$, where $H_0$ is the (local) inter-chain coupling between chains $A$ and $B$ and 
$H_{\rm pert}$ represents the Hamiltonian of the two decoupled chains. 
The ground state of the coupled system, to zeroth order in perturbation theory
is the product state $\big| \Psi_0\rangle=|1\rangle \times |2\rangle \times  \ldots \times |N\rangle$ where $\{ \big| n \rangle \}$ (with $n=1,\ldots N$ for chains of $N$ sites) 
are the states of the degrees of freedom of the 
two chains at the $n$th rung of this ladder. This ground state is non-degenerate and has a finite (and large) energy gap to all excitations. 
Since it is a product of singlet states, it is also ``maximally entangled'' even though in this basis the ground state is a product state. 
This is a simple example showing that the degree of entanglement of a state depends on how the question is posed, i.e. on the choice of the entangling region. 
Thus if we choose as the entangling region the left half of the ladder we would conclude that its entanglement entropy would be trivially zero. 
In contrast, if we choose one chain of the ladder as the entangling region the entanglement entropy will be (trivially) maximal).

We can compute next the corrections to the unperturbed ground state $\big|\Psi_0\rangle$ using an expansion in powers of the intra-chain interactions. 
Since we start with a gapped phase, the strong coupling expansion works well. 
For the sake of definiteness we will consider the problem of a  quantum Heisenberg antiferromagnetic model with $S=1/2$ on a two-leg ladder as an example. 
Other models can be treated using a similar procedure. 

The unperturbed Hamiltonian $H_0$ now contains only the inter-chain exchange interactions 
(with coupling constant $J_\perp$)  on the rungs of the ladder,
\begin{equation}
H_0=J_\perp \sum_{n=1}^N \vec S_{A}(n) \cdot \vec S_{B}(n)
\end{equation}
The coupling between the chains $A$ and $B$ is anti-ferromagnetic with $J_\perp>0$. 
For $H_0$, the ground state is the product of $N$ spin singlets on the rungs
\begin{equation}
\big|\Psi_0\rangle=\prod_{i=n}^N \big| 0,0\rangle_n
\label{eq:Psi0}
\end{equation}
where
\begin{equation}
\big| 0,0 \rangle_n=\frac{1}{\sqrt{2}} \big(\big|\uparrow_A,\downarrow_B\rangle_n-\big|\downarrow_A,\uparrow_B\rangle_n\big)
\end{equation}
is the spin singlet state on the $n$th rung of the ladder.
In the first excited state of the ladder, $\big|\Psi_1\rangle$, the spin singlet state of one rung is replaced by a spin triplet state $\big|1,m\rangle$, with $m=\pm 1, 0$ 
given by their standard expressions, $\big| 1,1\rangle=\big| \uparrow_A, \uparrow_B\rangle$, $\big|1,-1\rangle=\big|\downarrow_A, \downarrow_B\rangle$, and
$\big| 1,0 \rangle=(\big| \uparrow_A, \downarrow_B\rangle +\big|\downarrow_A, \uparrow_B\rangle)/\sqrt{2}$. The excitation energy is $E_1-E_0=J_\perp $. 
For the second excited state $\big|\Psi_2^{i,j}\rangle$, two singlets at rungs $i$ and $j$ become triplets, etc. 
For the $k$th excited state $\big|\Psi_k\rangle$, the excitation energy is $k J_\perp $.

The perturbing Hamiltonian $H_{\rm pert}$ is the sum of the Hamiltonians of the quantum Heisenberg antiferromagnets of the two chains
\begin{align}
H_{\rm pert}=&J \sum_{n=1}^N (\vec S_{A}(n) \cdot \vec S_{A}(n+1)+\vec S_{B}(n) \cdot \vec S_{B}(n+1)) \nonumber \\
 =&\frac{J}{2} \sum_{n=1}^N \left(\sigma^+_{A}(n)\sigma^-_{A}(n+1)+\sigma^-_{A}(n)\sigma^+_{A}(n+1)\right)\nonumber\\
  +&\frac{J}{4} \sum_{n=1}^N \sigma^z_{A}(n)\sigma^z_{A}(n+1)\nonumber \\
+&\frac{J}{2}\sum_{n=1}^N\left(\sigma^+_{B}(n)\sigma^-_{B}(n+1)+\sigma^-_{B}(n)\sigma^+_{B}(n+1)\right)\nonumber\\
+&\frac{J}{4}\sum_{n=1}^N\sigma^z_{B}(n)\sigma^z_{B}(n+1)
\label{eq:Hpert}
\end{align}
where we have expressed the spin operators in terms of the Pauli matrices.

Let us compute the ground state of the ladder to  first order in perturbation theory in $H_{\rm pert}$. By inspection of Eq.\eqref{eq:Hpert} we see that the only non vanishing 
contribution involves breaking the spin singlets on pairs of nearest-neighbor rungs at a time, i.e.  only $\langle \Psi^{n,n+1}_2\big|H_{\rm pert}\big|\psi_0\rangle \neq 0$. 
The perturbed ground state is
\begin{align}
\big|  \Psi \rangle = &\big|\Psi_0\rangle+\sum_{n}\frac{\langle\Psi^{n,n+1}_2\big|H_{\rm pert}\big|\Psi_0\rangle}{E_0-E_2}\big|\Psi^{n,n+1}_2\rangle \nonumber \\
=&\big|\Psi_0\rangle\nonumber\\
  -\sum_{n} & \left(-\frac{J}{16 J_\perp}\big|\phi_1^{n,n+1}\rangle-\frac{J}{16 J_\perp}\big|\phi_2^{n,n+1}\rangle+\frac{J}{4J_\perp}\big|\phi_3^{n,n+1}\rangle\right)\nonumber\\
&
\label{eq:perturbed-wf}
\end{align}
where $\{ \big|\phi_{1,2,3}^{n,n+1}\rangle \}$ are three different types of excited states of the unperturbed Hamiltonian $H_0$. 
In these excited states spins on pairs of nearest-neighbor rungs are put in triplet states. They are given by
\begin{align}
\big|\phi_1^{n,n+1}\rangle=&\ldots \big|1,1\rangle_n \big|1,-1\rangle_{n+1}\ldots \nonumber\\
\big|\phi_2^{n,n+1}\rangle=&\ldots \big|1,-1\rangle_n \big|1,1\rangle_{n+1} \ldots \nonumber\\
\big|\phi_3^{n,n+1}\rangle=&\ldots \big|1,0\rangle_n \big|1,0\rangle_{n+1}\ldots
\end{align}
where $\ldots$ represents product of singlets on the other rungs. We have
\begin{eqnarray}
\nonumber \langle \phi_{1}^{n,n+1}\big|H_{\rm pert}\big|\Psi_0\rangle&=&-J/8\\
\nonumber \langle \phi_{2}^{n,n+1}\big|H_{\rm pert}\big|\Psi_0\rangle&=&-J/8\\
 \langle \phi_{3}^{n,n+1}\big|H_{\rm pert}\big|\Psi_0\rangle&=&J/2
\end{eqnarray}

The wavefunction of Eq.\eqref{eq:perturbed-wf} is written in the basis of states of total spin state on the rungs. 
However in order to compute the reduced density matrix of one chain 
we will need to express the wave function in the basis of the spin projections of each chain, $ \big|S^z(1),\ldots,S^z(N)\rangle_A $ for chain $A$, and 
$ \big|S^z(1),\ldots,S^z(N)\rangle_B $ 
for chain $B$, respectively. Let us denote the spin configurations in chain $A$ by $\big|\phi\rangle_A$ and the spin configurations of chain $B$ by $\big|\phi\rangle_B$.

In this basis the unperturbed wave function $\big|\Psi_0\rangle$ of Eq.\eqref{eq:Psi0} is given by
\begin{align}
\big|\Psi_0\rangle=&\sum_{C(B)} \left(\frac{1}{\sqrt{2}}\right)^N (-1)^{m_d(C(B))} \big|\phi\rangle_A\big|\phi\rangle_B \nonumber\\
=\sum_{C(B)} &\left(\frac{1}{\sqrt{2}}\right)^N (-1)^{m_d(C(B))} \big|\uparrow\downarrow...\downarrow\uparrow...\rangle_A \; 
\big|\downarrow\uparrow...\downarrow\uparrow...\rangle_B
\label{eq:Psi0-product}
\end{align}
where we have denoted by $C(B)$ the set of all spin configurations in chain $B$, and by $m_d(C(B))$ the number of down spins $\downarrow$ in the configuration of chain 
$B$. Notice that in this basis  the spin configurations $\big| \phi\rangle_A$ of chain $A$ are  antiparallel to the spin configurations $\big| \phi\rangle_B$ in chain $B$ at every rung of the ladder.
Although this is a product state, this state is maximally entangled when the cut is made between the chains.

Similarly, when we add $H_{\rm pert}$ to $H_0$, in the basis of the spin projection of each chain, the perturbed wavefunction $\big|\Psi\rangle$ defined in Eq.\eqref{eq:perturbed-wf} can be rewritten in the following form:
\begin{widetext}
\begin{eqnarray}
\nonumber \big|\Psi\rangle &=&\sum_{C(B)}  \left(\frac{1}{\sqrt{2}}\right)^N(-1)^{m_d(C(B))}
 \left[(1-\frac{J }{4J_\perp}M_1+\frac{J }{4J_\perp}M_2)
\big|\phi\rangle_A -\sum_{C(A)^{\prime}}\frac{J}{8J_\perp} \big|\phi^{\prime}\rangle_A\right] 
\big|\phi\rangle_B\\
\nonumber &=&\sum_{C(B)}  \left(\frac{1}{\sqrt{2}}\right)^N(-1)^{m_d(C(B))}
 \left[(1-\frac{J }{4J_\perp}M_1+\frac{J }{4J_\perp}M_2)
\big|\uparrow...\uparrow\downarrow...\rangle_A -\sum_{C(A)^{\prime}}\frac{J}{8J_\perp} \big|\uparrow...\downarrow\uparrow...\rangle_A\right] 
\big|\downarrow...\downarrow\uparrow...\rangle_B\\
\label{eq:perturbedwf}
\end{eqnarray}
\end{widetext}
where $C(B)$ represents all the spin configurations $\big|\phi\rangle_B$ of chain $B$ (which are presented schematically in Eq.\eqref{eq:perturbedwf}). For each $\big|\phi\rangle_B$, the spin configuration $\big|\phi\rangle_A$ of chain $A$ is antiparallel with the spin configuration in chain $B$. $\big|\phi^{\prime}\rangle_A$ is defined by flipping the neighboring antiparallel spin paris ($\uparrow\downarrow$ or $\downarrow\uparrow$) in $\big|\phi\rangle_A$ and $C(A)^{\prime}$ represents all possible spin configurations for $\big|\phi^{\prime}\rangle_A$.
$m_d(C(B))$ is the number of down spins $\downarrow$ in the states of the $B$ chain, $M_1$ and $M_2$ are the numbers of pairs for parallel spins 
($\uparrow\uparrow$ or $\downarrow\downarrow$) and antiparallel spins ($\uparrow\downarrow$ or $\downarrow\uparrow$) in $\big|\phi\rangle_A$.

 To get the reduced density matrix for chain $A$, 
we need to use the Schmidt decomposition to trace out the states in chain $B$. The resulting (unnormalized) reduced density matrix for chain $A$ is
\begin{widetext}
\begin{align}
\rho_A=&\sum_{C(A)} \frac{1}{2^N}\left[\left(1-M_1 \frac{J}{2J_\perp}+M_2 \frac{J}{2J_\perp}\right) 
 \big|\phi\rangle_A \langle\phi\big|_A 
-\sum_{C(A)^{\prime}}\left( \frac{J}{4J_\perp} \big|\phi\rangle_A \langle\phi^{\prime} \big|_A+\textrm{h.c.}\right)\right]
\nonumber\\
=&\sum_{C(A)} \frac{1}{2^N}\left[\left(1-M_1 \frac{J}{2J_\perp}+M_2 \frac{J}{2J_\perp}\right) 
 \big|\uparrow...\downarrow\uparrow...\rangle_A \langle\uparrow...\downarrow\uparrow... \big|_A 
-\sum_{C(A)^{\prime}}\left(\frac{J}{4J_\perp} \big|\uparrow...\downarrow\uparrow...\rangle_A \langle\uparrow...\uparrow\downarrow... \big|_A+\textrm{h.c.}\right)\right]
\nonumber\\
&
\label{eq:rhoA-ladder}
\end{align}
\end{widetext}
where $C(A)$ are all the spin configurations in chain $A$ and $C(A)^{\prime}$ are the spin configurations obtained by flipping neighboring 
antiparallel spin pairs in $\big|\phi\rangle_A$.

The reduced density matrix for chain $A$ can be computed straightforwardly at this (first) order in perturbation theory in $J/J_\perp$.
 It has the form
 \begin{equation}
 \rho_A=
 \frac{1}{Z}  (1-\beta_{\rm eff} H_E+\ldots)\simeq \frac{1}{Z} e^{-\beta_{\rm eff} H_E+\ldots}
\label{eq:rhoA}
\end{equation}
where $Z$ normalizes the reduced density matrix, and
$H_E$ is the entanglement Hamiltonian. Notice that in Eq.\eqref{eq:rhoA}, in the square bracket, there are two terms, the first term 
$\big|\phi\rangle_A\langle\phi\big|_A$ can be understood as the potential term and the second term $\big|\phi\rangle_A\langle\phi^{\prime}\big|_A$ 
represents the hopping term between neighboring sites. Thus $H_E$ (at this order) is the Hamiltonian of the spin-1/2 antiferromagnetic quantum Heisenberg chain,
\begin{align}
\nonumber H_E=& \frac{J}{4} \sum_{n} \Big(2\sigma^z(n)\sigma^z(n+1)\nonumber \\
&+\sigma^+(n)\sigma^-(n+1)+\sigma^-(n)\sigma^+(n+1)\Big)+\ldots \nonumber \\
=& J \sum_{n}\vec S_{A}(n) \cdot \vec S_{A}(n+1)+\ldots
\label{eq:entanglement-H-highT}
\end{align}

Thus, in the strong coupling limit,  $J_\perp\gg J$, the reduced density matrix $\rho_A$ of chain $A$ is equal to the thermal density matrix 
$\rho_T$ of the chain with an effective (very high) temperature $T_{\rm eff}=2J_\perp\gg J$. In this limit the entanglement entropy equals to thermal entropy of the chain.  
  
The result we derived is a general consequence of the strong coupling limit and it is not peculiar to a ladder system. 
It is straightforward to see that, for instance, it also applies to a 2D bilayer antiferromagnet in the  regime of strong inter-layer exchange interactions. 
In this regime the bilayer system is gapped and the ground state is also well approximated by a product of singlets on the inter-layer couplings. 
By construction,  in all cases the resulting reduced density matrix  always describes a system at very high temperature.
 Thus we obtain that the reduced density matrix is thermal with an effective local Hamiltonian which that of a 2D quantum Heisenberg antiferromagnet. 
 Since the effective temperature is much larger than the intra-layer exchange interaction, the reduced density matrix of layer $A$ describes  
the paramagnetic phase of a single-layer  antiferromagnet.
However, this result does not imply that the entanglement Hamiltonian must necessarily always be equal to the Hamiltonian of the subsystem. 
For instance,  La\"uchli and Schliemann have also shown that at second order in perturbation theory the entanglement Hamiltonian acquires a 
 next-nearest-neighbor exchange interaction. Higher order terms in perturbation theory will generate more non-local terms in the effective Hamiltonian.

\section{Weak Coupling Limit}
\label{sec:weak-coupling}

From the discussion in Section \ref{sec:free-fermion} we  see that for the free fermion model, in the  strong tunneling limit, the reduced density matrix of one chain has thermal form, $\rho_{A}=\rho_T$.
However, in the same section we also saw that for the low energy modes of a chain of the ladder, i.e. those with wave vectors around $k=0$ and $k=\pi$, 
the reduced density matrix of one chain is also thermal regardless of the strength of the tunneling matrix element $t_\bot$. 
Also in section \ref{sec:strong-coupling} we saw that in the case of antiferromagnets on ladders, the reduced density matrix of one chain of the ladder 
is also thermal in the strong inter-ladder coupling, albeit with a temperature large compared with the scale of the entanglement Hamiltonian (which has the quantum Heisenberg form). 
By comparison with the results of Section \ref{sec:free-fermion} we would also expect that the reduced density matrix for the long-wavelength degrees of freedom of a chain 
of the ladder should also have a thermal form. This issue cannot be addressed by a direct calculation from the inter-chain strong-coupling regime of the ladder. 

In this section, we will consider the general case in the weak coupling limit.
We consider a system  with two critical chains with the same Hamiltonian which in the low-energy and long-wavelength limit describes a conformal field theory (CFT) in $1+1$ dimensions. 
We will further assume that, when coupled by some relevant operator $O(A,B)$ of the CFT, the combined system flows to a fixed point with a finite energy gap in its spectrum. 
Our goal is to determine if the reduced density matrix of one subsystem, $A$, has a thermal form.

Formally, the Hamiltonian of the coupled CFTs has the form
\begin{equation}
H=H_A+H_B+\int dx \; g\; O(A,B)
\label{eq:HAB}
\end{equation}
where $H_A\simeq H_B$  describe the two critical subsystems (the ``legs''), $O(A,B)$ is a suitable local relevant operator, and $g$ is a coupling constant. 
We will assume that this operator has the form $O(A,B)=\phi(A) \phi(B)$ where $\phi(A)$ and $\phi(B)$ are local operators of $A$ and $B$ each with (the same) scaling dimension 
$\Delta_{\phi(A)}=\Delta_{\phi(B)}\equiv \Delta/2$. This perturbation is relevant if its scaling dimension $\Delta_{\phi(A)}+\Delta_{\phi(B)}=\Delta\leq 2$ (where $2$ is the space-time dimension). 
Under these assumptions this perturbation drives the combined system into a massive phase with a finite mass gap $M(g)$ which obeys the scaling relation 
$M(g) \sim \textrm{const.} \; g^{\nu z}$ where $\nu=2-\Delta$. These CFTs are ``relativistic'' and hence have dynamical exponent  $z=1$. 
The case $\Delta=2$ is special in that the operator $O(A,B)$ is marginal. We will further assume that in this case it is marginally relevant. 

In the case of the fermionic ladder of Section \ref{sec:free-fermion} the CFT of the decoupled chains is a theory of two massless Dirac (Weyl) fermions (and hence with central charge $c=2$). 
The scaling dimension of the tunneling operator (i.e. the fermion mass term) is $\Delta=1$ which is relevant. In this case, the exponent is $\nu=1$. 
In the case of the two-leg ladder, the decoupled ladder is a theory of two spin-1/2 quantum Heisenberg antiferromagnetic chains and hence are critical. 
The CFT of the spin-1/2 quantum Heisenberg antiferromagnetic chain is an SU(2)$_1$ Wess-Zumino-Witten (WZW) model.\cite{Affleck1986b} 
Hence the decoupled ladder is a product of two SU(2)$_1$ WZW models  (with total central charge $c=2$). The most relevant operator in the inter-ladder exchange interaction is the coupling of the 
N\'eel order parameters of each chain, $\vec N_A(x) \cdot \vec N_B(x)$. In the SU(2)$_1$ CFT the N\'eel order parameters of each chain are represented by the primary field 
whose scaling dimension is $1/2$ (for a detailed discussion see, e.g. Ref.[\onlinecite{Fradkin1991}]). 
Hence, the scaling dimension of the inter-chain exchange interaction in the spin-1/2 ladder is $\Delta=1$, and hence the exponent is $\nu=1$ (albeit for different reasons than in the case of the fermionic ladder).

\begin{figure}[t]
\psfrag{tau}{$\tau$}
\begin{center}
\includegraphics[width=0.4\textwidth]{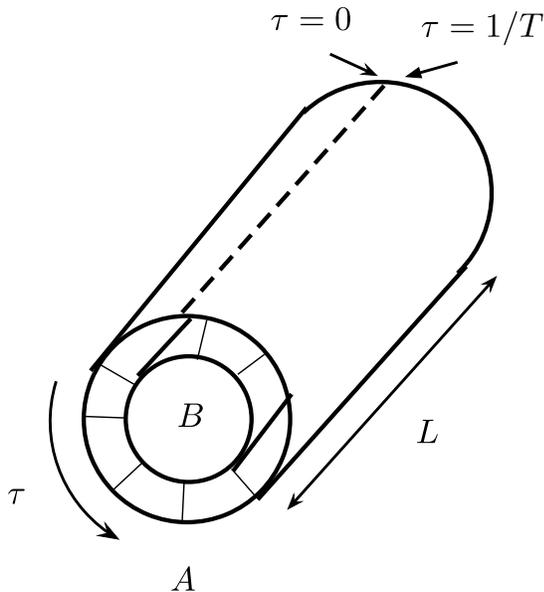}
\end{center}
\caption{Spacetime manifold with a cut required for the computation of $\rho_A$. 
The cut (the broken line) only affects the spacetime for subsystem $A$ (the outside cylinder) whose configurations are discontinuous across the cut. 
The configurations on region $B$ (the inside cylinder) are periodic and smooth. The interactions between the fields on regions $A$ and $B$ is depicted by 
the thin radial lines.}
\label{fig:rhoA}
\end{figure}

The computation of the reduced density matrix of a subsystem (in this case a perturbed CFT) is in general a very difficult problem. 
In principle it is possible to compute the reduced density matrix  using methods of quantum field theory which reduces this computation to an imaginary time  path integral over 
the field configurations $\phi(x,\tau)$, with $0\leq x \leq L$ and $0\leq \tau\leq 1/T$ (in the limits $L \to \infty$ and $1/T \to \infty$), with suitable boundary conditions. 
For the matrix element $\langle \phi_A^{\textrm{in}}(x)\big|\rho_A\big|\phi^{\textrm{out}}_A(x)\rangle$, 
the boundary conditions are that the field configurations for region 
$B$ are periodic in imaginary time, $\phi_B(x,0)=\phi_B(x,1/T)$, whereas on region $A$ the field configurations 
are discontinuous across the $x$ axis between $\tau=0$ and $\tau=1/T$, and hence satisfy $\phi_A(x,0)=\phi_A^{\textrm{in}}(x)$ and 
$\phi_A(x,1/T)=\phi_A^{\textrm{out}}(x)$ (see Ref. [\onlinecite{Calabrese2004}]). 
For the type of problems we are discussing here the result is a path integral on two concentric cylinders each of length $L$ and and circumference $1/T$, 
with the cylinder for region $A$ having a cut along the $x$ axis representing the discontinuity of the field configurations, as shown in Fig.\ref{fig:rhoA}.

\begin{figure}[t]
\begin{center}
\includegraphics[width=0.4\textwidth]{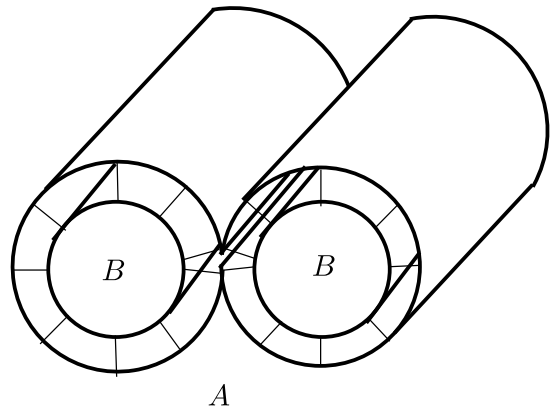}
\end{center}
\caption{ The spacetime manifold needed for the computation of $\textrm{Tr}\rho_A^2$. The inside cylinders represent the replicated regions $B$ 
(which are integrated out) and the outer surface which wraps around them is the replicated $A$ region. 
The interactions between the $A$ and $B$ regions are shown as thin radial lines.}
\label{fig:trrhoA^2}
\end{figure}

Alternatively, we can compute the moments of the reduced density matrix of the subsystem (needed for the computation of the R\'enyi and von Neumann entropies) 
using the replica trick\cite{Callan1994,Holzhey1994,Calabrese2004}
\begin{equation}
\textrm{Tr}\rho_A^n=\frac{\mathcal{Z}_n}{\mathcal{Z}^n}
\label{eq:Z_n/Z^n}
\end{equation}
from which the R\'enyi entropies $S_n$ and the von Neumann entropy can be determined,
\begin{equation}
S_n=\frac{1}{1-n} \ln \textrm{Tr} \rho^n_A, \qquad S_{vN}=\lim_{n\to 1} S_n
\end{equation}
In Eq.\eqref{eq:Z_n/Z^n} we have denoted by  $\mathcal{Z}$  the partition function of the coupled system (with coupling constant $g$) defined on a cylinder of length $L \to \infty$ 
and circumference $1/T \to \infty$. $\mathcal{Z}_n$ is the partition function of the coupled system (with subsystems $A$ and $B$) 
on a spacetime manifold obtained by stitching together $n$ copies of the path integral of the reduced density matrix. In the case at hand this leads to the manifold  
shown in Fig.\ref{fig:trrhoA^2} (for the case $n=2$), where the $B$ region are the inside cylinders whereas the $A$ region is obtained by gluing together the $n$ 
path integrals along the $n$ cuts. Therefore, $\mathcal{Z}_n$ is a path integral in which the fields on the $n$ copies of the region $B$ are periodic with period $1/T$.
Instead the fields on region $A$ are stitched together in such a way that they are periodic with period $n/T$ (see Fig. \ref{fig:trrhoA^2}). 
The partition function $\mathcal{Z}$ should not be confused withe the normalization $Z$ of the reduced density matrix.


This procedure requires the introduction of a set of twist fields that connect the Hilbert spaces two at a time. 
In the case of spatial cuts there are a finite number of such twist fields. In the case of a conformally invariant theory the twist fields behave as local operators with 
non-trivial scaling dimensions and uniquely determine the singularities of the path-integral.\cite{Calabrese2004} However in the case in which two conformal field 
theories (on regions $A$ and $B$) are coupled everywhere we are led to the ``body'' cuts we described above (and shown in Fig.\ref{fig:rhoA}) 
which require the introduction of a line of twist fields defined along these cuts.

The introduction of this line of twist fields complicates the calculation of the replicated partition function, and we will not pursue this approach here. Another option is to use 
the approach introduced by Qi, Katsura and Ludwig\cite{Qi2012} who made the observation that upon physically splitting regions $A$ and $B$ suddenly, 
i.e. upon setting the coupling constant $g\to 0$ after some (real) time $t=0$, the reduced density matrix of subsystem $A$ becomes the density matrix of the 
(now decoupled) system $A$. These authors used this approach to relate the entanglement entropy of a simply connected region of a 
2D chiral topological phase to the behavior of its edge states.

In this section we will formulate instead a scaling argument to generalize the results of Section \ref{sec:free-fermion}. There we saw that the reduced density matrix 
of the long-wavelength modes of a  chain of a gapped  free fermion system on a ladder is thermal and that the von Neumann entropy of the chain is the thermal entropy 
of an isolated chain at a finite effective temperature set by the gap in the fermion spectrum. We also saw that the resulting expressions for the entanglement entropies 
(von Neumann and R\'enyi)
depend only on the Casimir term that gives the form of the finite size correction to the free energy in a conformal field theory. 
The structure of the universal Casimir term is determined by conformal invariance and by the conformal anomaly\cite{Blote1986,Affleck1986} 
(through the central charge $c$). 
We are thus led to conjecture that this behavior of the entanglement entropies holds for any system of two coupled conformal field theories in  
a massive phase with a mass gap $M(g) \sim g^\nu$.

The scaling argument 
is based on the observation that the quantity $\mathcal{F}_n=-T \ln \mathcal{Z}_n$ is the {\em free energy} of the 
replicated system and, as such, it is a function of $L$, $T$ and $n$ (as well as of the coupling constant $g$). 
The scaling behavior is expected to hold since we are dealing with a perturbed conformal field theory which, due to 
the effects of the relevant 
perturbation, is driven into a massive phase. 
Since the coupled theory now has a finite mass gap $M(g)$ and a finite correlation length $\xi(g)$, the singular part of 
$\ln \mathcal{Z}$ of the coupled system, whose Hamiltonian is given in Eq.\eqref{eq:HAB}, should be, as in all theories of critical behavior,\cite{cardy-book} 
an extensive homogeneous function of the form (known in the theory of Critical Phenomena as Widom scaling)
\begin{equation}
\big(\ln \mathcal{Z}\big)_{\rm sing}=  \textrm{const.} \; T L\; \xi^{-2}(g)\; f(g)
\label{eq:scaling-lnZ}
\end{equation}
where $f(g)$ is a function such that $f(0)=1$. 

Turning now to the replicated partition function, $\mathcal{Z}_n$, we notice that on the $A$ region the stitched cuts act only at imaginary times 
$\tau=p/T$ (with $p=1,\ldots,n$) and for all values of $x$. 
The partition function of the replicated system, $\mathcal{Z}_n$, differs from the partition function of a single copy by the action of the lines of twist fields at $n$ 
equally spaced boundaries in imaginary time. 
We are interested in the limit in which both $L \to \infty$ and $T \to \infty$ for fixed and finite $n$. 
In this limit $\mathcal{Z}_n$ should have a bulk contribution which is asymptotically the same as the bulk contribution of $n$ decoupled copies. 

By examining the free energies $-T \ln \mathcal{Z}_n$ and $-nT \ln \mathcal{Z}$, we notice that in the thermodynamic limit $L \to \infty$ and $T \to 0$, 
the bulk contributions should cancel exactly each other out and that the only surviving contributions come from the ``defects'' (associated with the twist fields). 
Thus, the piece we are interested in is a finite size correction in $\mathcal{Z}_n$ which defines a type of boundary field theory. 
Furthermore, since for any finite value of the coupling constant $g$ 
the theory is in a massive phase, the subtracted quantity $(\ln \mathcal{Z}_n- n\ln \mathcal{Z})$ (needed to compute the R\'enyi entropies) 
has contributions only from a strip of width $\xi=1/M(g)$ and length $L$. 
The length scale $\xi$ is the ``extrapolation length'' invoked in Refs.[\onlinecite{Gambassi2011,Qi2012}]. 

Therefore, again in the thermodynamic limit $L \to \infty$ and  $T \to 0$, we expect to obtain the scaling behavior
\begin{equation}
\lim_{T\to 0, L \to \infty} \big(\ln \mathcal{Z}_n-\ln \mathcal{Z}^n\big)= L M(g)\; \tilde f_n(g)
\label{eq:sing-Renyi}
\end{equation}
where $\tilde f_n(g)$ is another function with the limit $\widetilde f_n(0)=f_n$. By demanding consistency with the results from Section \ref{sec:free-fermion}, 
we will conjecture that the quantities $f_n$ are given by
\begin{equation}
f_n=\frac{\pi c}{6v} \left(\frac{1}{n}-n\right) 
\label{eq:fn}
\end{equation}
where $c$ is the central charge of each of the two conformal field theories at $g=0$ and $v$ is the velocity of their long-wavelength modes. 

We are then led to conjecture that the von Neumann entanglement entropy of subsystem $A$ of a gapped system $A\cup B$ is extensive and has the scaling behavior
\begin{equation}
S_{vN}=\frac{\pi c}{3v} M(g) L
\end{equation}
where $M(g) \sim g^{2-\Delta}$, $c$ is the central charge $c$ of the decoupled CFTs which are coupled by a local relevant operator of scaling dimension $\Delta$, and $v$ is the velocity of the modes. These arguments also imply that the R\'enyi entropies $S_n$ should be given by an expression of the form
\begin{equation}
S_n=\frac{\pi}{6} \frac{c}{v} \left(\frac{1}{n}+1\right) M(g) L
\end{equation}

\section{Gapless Coupled Luttinger Liquids}
\label{sec:coupled-LL}

For completeness, in this Section we will consider a situation where the coupling operator $O(A, B)$ is marginal and therefore will not open a gap in the spectrum. 
The entanglement entropy thus should be different from the thermal entropy. As a simple example we consider two Luttinger liquids coupled with a marginal operator. 
The R\'enyi entropy for this model has been calculated  before by Furukawa and Kim, using the replica trick. They  showed that the von Neumann entanglement entropy has, 
in addition to a term proportional to the length of the subsystem, there is a constant term determined by Luttinger parameter.\cite{Furukawa2011} 
Here we will arrive to the same result using a different (and simpler) method. We will obtain this result directly by computing the reduced density matrix  $\rho_A$. 
This can be done since the Luttinger liquid model is essentially a free scalar (compactified) (Bose) field.

The Hamiltonian  density for this model is
\begin{equation}
\mathcal{H}=\mathcal{H}_A+\mathcal{H}_B+\mathcal{H}_{AB}
\label{eq:HLutt-total}
\end{equation}
where $\mathcal{H}_A$ and $\mathcal{H}_B$ are the Hamiltonian densities for the two Luttinger liquids
\begin{equation}
\mathcal{H}_{AB}=\frac{v}{2} \left[\frac{\Pi^2}{K}+K(\partial_x \phi)^2\right]
\label{eq:HLutt}
\end{equation}
In momentum space the Hamiltonians have the form
\begin{equation}
H_{AB}= \sum_{p \neq 0}v|p| \left(a^{\dag}_p a_p+\frac{1}{2}\right)+\frac{v}{2LR^2}M^2+\frac{2 v}{L}R^2N^2
\label{eq:HLutt-momentum}
\end{equation}
where $\phi$ is a compactified boson with compactification radius $r$ and $R=r \sqrt{K}=1/\sqrt{4\pi}$, where $K$ is the Luttinger parameter, and $\Pi$ is the canonical momentum conjugate to the the  field $\phi$. 
Here $M$ and $N$ take integer values. (For a summary 
of the Luttinger model see, e.g., Refs. [\onlinecite{Fradkin1991}] and [\onlinecite{Gogolin1998}]).

The coupling term $\mathcal{H}_{AB}$ takes the form
\begin{equation}
\mathcal{H}_{AB}=  uK \partial_x \phi_A \partial_x \phi_B - \frac{u}{K} \Pi_A\Pi_B
\label{eq:HABLutt}
\end{equation}
where $u$ is the coupling constant.
In momentum space the inter-chain coupling  Hamiltonian is 
\begin{align}
H_{AB}=&\sum_{p\neq 0} u|p|(a_{p}^{\dag}b^{\dag}_{-p}+a_pb_{-p})\nonumber\\
-&\frac{u}{LR^2}M_AM_B+\frac{4 u}{L}R^2N_AN_B
\label{eq:HLuttAB}
\end{align}
where $a_p$ and $b_p$ are the boson operators for chain $A$ and chain $B$, respectively.

The inter-chain coupling term of Eq.\eqref{eq:HABLutt}, has scaling dimension $2$ and hence it is a marginal operator. In the case of the Luttinger model it is an exactly marginal operator.
Its main effects are to change (continuously) the scaling dimensions of the operators of the physical observables, as well as a finite renormalization of the velocities of the modes 
(see, e.g., Refs.[\onlinecite{Vishwanath2001}] and [\onlinecite{Emery2000}]). The coupled Luttinger models are stable provided $|u|<v$.

Since $|u|<v$, the ground state of this system is in the sector where the winding modes are absent, $N_A=N_B=M_A=M_B=0$. 
Thus, we only need to solve the following Hamiltonian:
\begin{equation}
H=\sum_{p\neq 0} \Big[ v|p|(a^{\dag}_p a_p+b^{\dag}_p b_p)+u|p|(a_pb_{-p}+a_p^{\dag}b_{-p}^{\dag})\Big]
\end{equation}
which is a bilinear form in the bosons. Since the number of bosons in the separate chains are not conserved, 
the diagonalization of the Hamiltonian then proceeds through the standard Bogoliubov transformation
\begin{align}
a_p^{\dag}=&f_+c_p^{\dag}+f_-d_{-p}\nonumber\\
 b_{-p}^\dagger=&f_+ d_{-p}^\dagger+f_- c_{p}
 \label{eq:Bogoliubov}
\end{align}
By diagonalizing the Hamiltonian, we can get the new spectrum for the bosons
\begin{equation}
E(p)=|p|\sqrt{v^2-u^2}
\end{equation}
The parameters $f_\pm$  are given by
\begin{equation}
 f_\pm^2=\frac{1}{2}\left((1-u^2/v^2)^{-1/2} \pm 1\right)
\end{equation}

Since the coupled Luttinger model has been reduced to a free bosonic model, the reduced density matrix for chain $A$ can be calculated similarly 
as in free fermionic model. 
The entanglement Hamiltonian here too has the form $\widetilde H_E=\sum_{ij}\widetilde H_{ij}a^{\dag}_ia_j$ with
\begin{equation}
\widetilde H_{ij}=\Big(\ln [C^{-1}-1]\Big)_{ij}
\end{equation}
where $C_{ij}$ is the correlation matrix. Its matrix elements in momentum space  (and in the thermodynamic limit $L \to \infty$) are
\begin{equation}
C_{pp^{\prime}}=2\pi \delta(p-p^{\prime})f_-^2
\end{equation}
Since $f_-^2$ is a constant, the matrix $\widetilde H_{ij}$ is proportional to the identity matrix. 
Hence the entanglement Hamiltonian is proportional to the number operator and it is not equal to the Hamiltonian of one of the subsystems. 
Consequently the reduced density matrix is no longer thermal. 

This difference is also reflected in the different behavior of the von Neumann entanglement entropy $S_{vN}$ and thermal entropy $S_T$. Let us define the parameter $\kappa$,
\begin{equation}
\kappa=\frac{K_{+}-K_{-}}{K_{+}+K_{-}}=\frac{u}{v}
\end{equation}
where 
\begin{equation}
K_\pm=K \left(\frac{v\pm u}{v\mp u}\right)^{1/2}
\label{eq:Kpm}
\end{equation}
are the Luttinger parameters for the fields $\phi_\pm=(\phi_A\pm \phi_B)/\sqrt{2}$ that diagonalize  the Hamiltonian of the coupled system, Eq.\eqref{eq:HLutt-total}. 
We will now obtain the expressions of the entanglement entropies as functions of $\kappa$.

In the weak coupling limit $|u|\ll v$ (i.e. $\kappa\ll 1$) and in momentum space, the correlation matrix is
\begin{equation}
C_{pp^{\prime}}
\simeq \frac{\kappa^2}{4} 2\pi \delta(p-p^{\prime})
\end{equation}
It follows that the R\'enyi entropies $S_n$ are equal to
\begin{eqnarray}
\nonumber S_n&=&\frac{1}{1-n}\ln \textrm{Tr}\rho_A^n\\
\nonumber &=&\frac{1}{1-n}\left(\frac{L}{a}-1\right)\ln \left[\frac{(1-e^{-E})^n}{1-e^{-nE}}\right]\\
\nonumber &\approx&\frac{1}{1-n}\left(\frac{L}{a}-1\right)\left(-n\frac{\kappa^2}{4}+\left(\frac{\kappa}{2}\right)^{2n}\right)\\
&=&-\gamma_n \frac{L}{a}+\gamma_n
\end{eqnarray}
where $E=\ln\left((4/\kappa^2)-1\right)$, $a$ is a short-distance cutoff and 
\begin{equation}
\gamma_n=\frac{1}{(1-n)}\left[n\frac{\kappa^2}{4}-\left(\frac{\kappa}{2}\right)^{2n}\right]
\end{equation}
 From the above equation, we see that besides a term proportional to the length $L$ of the system, there is also a constant term related to the Luttinger liquid parameter. 
 When $n$ is large, $\gamma_n=\frac{n\kappa^2}{4(1-n)}$. These results  agree with those of Ref.[\onlinecite{Furukawa2011}]. 
 Similarly, the von Neumann entanglement entropy equals to 
\begin{equation}
S_{vN}=\left(\frac{L}{a}-1\right)\frac{\kappa^2}{4}\left[1-\ln \left(\frac{\kappa^2}{4}\right)\right]=-\gamma_1 \frac{L}{a}+\gamma_1
\end{equation} 
where $\gamma_1=-\frac{\kappa^2}{4}(1-\ln\frac{\kappa^2}{4})$. We can see that the von Neumann entanglement entropy  $S_{vN}$ for this system  is extensive but it is 
totally different from the thermal entropy $S_T$ which is given by Eq.\eqref{eq:1D-cft-entropy} (with $c=1$).

\section{Conclusions}
\label{sec:conclusions}

In conclusion, in this work we obtained the reduced density matrix in some two-leg ladder systems. 
We find that when the two chains that are critical and are coupled by some relevant operator which opens a finite energy gap in the spectrum,  
the reduced density matrix for one chain takes the same form as the thermal density matrix with the energy gap playing the role of the effective temperature. 
This idea is verified at both the strong coupling limit and the weak coupling limits. We also noted that although the entanglement Hamiltonian is generally non-local, 
the reduced density matrix for the long-wavelength modes of the subsystem is of the Gibbs form with a local effective Hamiltonian  with a finite effective temperature. 
The fraction of modes which are thermal increases as the strength of the coupling increases. 
We showed that the entanglement von Neumann entropy for the long wavelength modes has a universal form which is equal to the thermodynamic entropy of 
the decoupled conformal field theory with central charge $c$. 
We verified the validity of this conjecture by explicit calculations in  a ladder fermionic system with a gap. 
The strong coupling results are generally valid and also hold in higher dimensional systems

\begin{acknowledgments}
We thank P. Calabrese, P. Fendley, S. Kivelson, A. La\"uchli, I. Peschel and D. Poilblanc for illuminating discussions. 
We also thank H. Katsura and R. Lundgren for correspondence and for alerting us about Refs. [\onlinecite{Katsura2010,Lou2011,Lundgren2012}], repsectively.
This work was supported in part by the National Science Foundation, under grant DMR-1064319 at the University of Illinois.
\end{acknowledgments}

\bibliographystyle{apsrev}
\bibliography{biblio}
\end{document}